\documentclass[prl,twocolumn,amsmath,amssymb,superscriptaddress]{revtex4}

\usepackage[pdftex]{graphics}
\usepackage{graphicx}
\usepackage{color}
\usepackage[T1]{fontenc}
\usepackage{ulem}

\begin{document}

\title{Charge puddles in the bulk and on the surface of the topological insulator BiSbTeSe$_2$ 
	studied by scanning tunneling microscopy and optical spectroscopy}

\author{T. Knispel}
\affiliation{II. Physikalisches Institut, Universit\"{a}t zu K\"{o}ln, Z\"{u}lpicher Strasse 77, D-50937 K\"{o}ln, Germany}
\author{W. Jolie}
\affiliation{II. Physikalisches Institut, Universit\"{a}t zu K\"{o}ln, Z\"{u}lpicher Strasse 77, D-50937 K\"{o}ln, Germany}
\affiliation{Institut f\"{u}r Materialphysik, Westf\"{a}lische Wilhelms-Universit\"{a}t M\"{u}nster, Wilhelm-Klemm-Strasse 10, 
	D-48149 M\"{u}nster, Germany}
\author{N. Borgwardt}
\affiliation{II. Physikalisches Institut, Universit\"{a}t zu K\"{o}ln, Z\"{u}lpicher Strasse 77, D-50937 K\"{o}ln, Germany}
\author{J. Lux}
\affiliation{Institut f{\"u}r Theoretische Physik, Universit\"{a}t zu K\"{o}ln, Z\"{u}lpicher Strasse 77, D-50937 K\"{o}ln, Germany}
\author{Zhiwei Wang}
\affiliation{II. Physikalisches Institut, Universit\"{a}t zu K\"{o}ln, Z\"{u}lpicher Strasse 77, D-50937 K\"{o}ln, Germany}
\author{Yoichi Ando}
\affiliation{II. Physikalisches Institut, Universit\"{a}t zu K\"{o}ln, Z\"{u}lpicher Strasse 77, D-50937 K\"{o}ln, Germany}
\author{A. Rosch}
\affiliation{Institut f{\"u}r Theoretische Physik, Universit\"{a}t zu K\"{o}ln, Z\"{u}lpicher Strasse 77, D-50937 K\"{o}ln, Germany}
\author{T. Michely}
\author{M. Gr\"{u}ninger\footnote{grueninger@ph2.uni-koeln.de}}
\affiliation{II. Physikalisches Institut, Universit\"{a}t zu K\"{o}ln, Z\"{u}lpicher Strasse 77, D-50937 K\"{o}ln, Germany}

\date{August 29, 2017}

\begin{abstract}
The topological insulator BiSbTeSe$_2$ corresponds to a compensated semiconductor in which strong Coulomb disorder gives rise 
to the formation of charge puddles, i.e., local accumulations of charge carriers, both in the bulk and on the surface. 
Bulk puddles are formed if the fluctuations of the Coulomb potential are as large as half of the band gap. 
The gapless surface, in contrast, is sensitive to small fluctuations but the potential is strongly suppressed 
due to the additional screening channel provided by metallic surface carriers. 
To study the quantitative relationship between the properties of bulk puddles and surface puddles, 
we performed infrared transmittance measurements as well as scanning tunneling microscopy measurements on the same sample 
of BiSbTeSe$_2$, which is close to perfect compensation. 
At 5.5\,K, we find surface potential fluctuations occurring on a length scale $r_s$\,=\,$40-50$\,nm 
with amplitude $\Gamma$\,=\,$8-14$\,meV which is much smaller than in the bulk, where optical measurements 
detect the formation of bulk puddles. 
In this nominally undoped compound, the value of $\Gamma$ is smaller than expected for pure screening by 
surface carriers, and we argue that this arises most likely from a cooperative effect of bulk screening and surface screening.
\end{abstract}

\maketitle
\section{Introduction}

Three-dimensional topological insulators (TIs) are narrow-gap semiconductors with band inversion 
\cite{Ando13}. Many TIs such as Bi$_2$Te$_3$ and Bi$_2$Se$_3$ show metallic bulk conductivity 
due to defect-induced charge carrier densities of typically 10$^{19}$\,cm$^{-3}$ 
\cite{Stordeur92,Analytis10,Ren10,LaForge10,Butch10,Eto10,Post13}. 
For investigations of the topological surface states, bulk-insulating samples are highly desirable. 
Bulk-insulating behavior was obtained in Bi$_{2-x}$Sb$_x$Te$_{3-y}$Se$_y$ \cite{Taskin11,Ren11} 
by compensation of donors and acceptors \cite{Ando13}. 
For perfect compensation, donor density $N_D$ and acceptor density $N_A$ are equal, 
$K \! \equiv \! N_A/N_D$\,=\,$1$, 
and electrons are transferred from donors to acceptors, suppressing the defect-induced charge density 
in both, conduction and valence bands. 
However, the randomly distributed ionized donors and acceptors give rise to Coulomb disorder, 
as discussed by Shklovskii and coworkers \cite{Shklovskii72,Skinner12,Skinner13,Chen13}. 
This yields strong band bending with potential fluctuations as large as $\Delta/2$, 
where $\Delta$ denotes the band gap. Accordingly, the bands locally touch and cross the chemical 
potential \cite{Skinner12}, giving rise to local accumulations of charge carriers, the so-called puddles. 
These puddles can be viewed as regions which are either $p$- or $n$-doped, both occurring with equal 
probability for $K$\,=\,$1$. Puddles partially screen the large fluctuations of the 
Coulomb potential, and they ``evaporate'' with increasing temperature due to the 
additional screening contribution of thermally activated carriers \cite{Borgwardt16}. 
For bulk transport, puddles are assumed to explain the small activation energy observed in the 
resistivity \cite{Skinner12}, the coexistence of electron-type and hole-type carriers in 
compensated samples \cite{Rischau16}, and the gigantic negative magnetoresistance observed in 
TlBi$_{0.15}$Sb$_{0.85}$Te$_2$ \cite{Breunig17}. 
However, the smoking gun for bulk puddles is a Drude-like feature in the optical conductivity 
which is suppressed on a temperature scale $k_B T$\,=\,$E_C$ given by the Coulomb interaction between 
neighboring dopants \cite{Borgwardt16}. 
Infrared transmittance data of BiSbTeSe$_2$ yield $E_C/k_B$\,=\,$30-40$\,K \cite{Borgwardt16}.

It is important to understand in how far these strong fluctuations of the bulk Coulomb potential 
and the concomitant dramatic band bending affect the properties of the topological surface state. 
Even minute potential fluctuations cause spatial variations of the Fermi level 
within the gapless Dirac cone and thus create local accumulations of charge carriers at the surface, 
i.e., surface puddles. Compared to the bulk, the Dirac-like surface state thus gives rise to 
strongly enhanced screening \cite{Skinner13}. 
However, for a nominally undoped TI with the Dirac point at the Fermi level yielding a vanishing 
density of states, a simple linear response theory predicts the absence of screening. 
Nonlinear screening is, however, effective \cite{Skinner13}
resulting in a reduced but finite amplitude of fluctuations of the surface potential. 
The corresponding spatial variations of the Fermi level are important for surface transport 
phenomena \cite{Skinner13,Xiong12a}. This is well established in graphene with 
Coulomb disorder in the substrate \cite{Hwang07,Adam07,Martin08,Zhang09,Samaddar16}. 
Spatial variations of the surface density of states are measurable by 
scanning tunneling spectroscopy (STS) in the differential conductance $d I / d V$. 
Indeed, STS data of doped Bi$_2$Te$_3$ and Bi$_2$Se$_3$ show a scatter in the Dirac point energy $E_D$ 
of $20-40$\,meV \cite{Beidenkopf11}. Also, spatial fluctuations of $E_D$ have been reported for 
Bi$_{1.5}$Sb$_{0.5}$Te$_{1.7}$Se$_{1.3}$ \cite{Kim14,Ko16} and were argued to be the source of 
scattering of long-wavelength electrons, leading to pronounced quasi-particle interference \cite{Kim14}.
However, a quantitative comparison of theoretical predictions and experimental results on 
the properties of both, surface puddles and bulk puddles measured in the same sample, e.g.\ by 
STS and optics, is still lacking.

The properties of bulk puddles and surface puddles -- such as the carrier density in bulk puddles, 
the temperature scale of bulk puddle destruction, the length scale of surface puddle formation, and 
the fluctuations of the surface potential -- are intimately linked to the degree of compensation $K$ 
and the Coulomb interaction between neighboring dopants, 
\begin{equation}
E_C = \frac{e^2}{4 \pi \epsilon_0 \epsilon}\, N_{\rm def}^{1/3} \, ,
\label{ec}
\end{equation}
with the elementary charge $e$, the dielectric constant $\varepsilon$, 
and the defect density $N_{\rm def}$\,=\,$(N_A+N_D)/2$ \cite{Skinner13,Borgwardt16}. 
Both $N_{\rm def}$ and $K$ are sample dependent. A systematic investigation of the relationship between 
bulk puddles and surface puddles as well as a quantitative comparison with theory therefore require 
to study bulk and surface properties on the same sample. 
The compound BiSbTeSe$_2$ is ideally suited for this task since it is close to perfect compensation \cite{Borgwardt16}. 
Moreover, angle-resolved photo-electron spectroscopy (ARPES) has shown that the Dirac point is close to the Fermi level 
in BiSbTeSe$_2$ \cite{Arakane12,Neupane12}, which enhances the sensitivity of the Fermi level to potential fluctuations. 
Here, we combine optical spectroscopy of BiSbTeSe$_2$ to study bulk puddles and scanning tunneling microscopy (STM) 
and STS to investigate charge puddles at the surface of the same sample. 
At low temperatures, we indeed observe the coexistence of bulk puddles and surface puddles. 
In contrast to expectations, our results indicate that a quantitative description of the potential 
fluctuations on the surface, i.e.\ of surface puddles, requires to take into account the screening contribution 
of bulk carriers. This can be rationalized by comparing the relevant length scales.

\section{Experimental methods}

The single crystals of BiSbTeSe$_2$ used in the present study were grown from high-purity elements [Bi, Sb, and Te of 6N (99.9999\%) 
and Se of 5N (99.999\%) purities] by using a modified Bridgman method in a sealed quartz-glass tube as described in Ref.\ \cite{Ren11}. 
To facilitate the \textit{in-situ} cleaving under UHV, the crystals were precut into platelets with a typical dimension of 
$3\times3$\,mm$^2$ with the (111) plane as the wide face.

For STM measurements two different BiSbTeSe$_2$ crystals were mounted with their backside to the STM sample holder by using 
silver-filled epoxy glue. On their topside a metal pin was attached normal to the crystal surface using the same glue. 
Cleaving was performed in the STM ultrahigh vacuum preparation chamber with $p<2 \times 10^{-10}$\,mbar by moving the pin 
against a sharp edge, causing the crystal to cleave. The cleaved sample surface was moved into the STM bath cryostat within 
a few minutes, where the pressure was $p<10^{-11}$\,mbar. The two crystals displayed indistinguishable results, 
underlining the reproducibility of our experiments.

STM and STS were performed at 5.5\,K and 77\,K.\@ Constant-height d$I$/d$V$ point spectra as well as constant-current d$I$/d$V$ maps 
were recorded, with $V$ being the bias voltage applied to the sample and $I$ the tunneling current. We refer to them as STS spectra 
and STS maps in the following. Both were measured with a lock-in amplifier using a modulation voltage $V_{\rm{mod}}$ of $4-10$\,mV 
and a frequency of $f$\,=\,$777$\,Hz. Each STS spectrum was averaged at least three times. The STS maps and STM topographs 
were analyzed with the WSxM software \cite{Horcas03}. 

Prior to the STM/STS investigations, we performed infrared transmittance measurements in the frequency range of 0.07\,eV to 
about 0.9\,eV using a Bruker IFS 66v/S Fourier-transform spectrometer equipped with a continuous-flow He cryostat. 
The data were recorded using unpolarized light with the electric field parallel to the cleavage plane. 
At 5\,K, the samples are transparent up to about 250\,meV, which corresponds to the band gap. 
The optical conductivity $\sigma_1(\omega)$ was calculated from the measured transmittance in combination with the 
reflectivity of a thick sample of BiSbTeSe$_2$, as described in Ref.\ \cite{Borgwardt16}.

\section{Experimental results}

\subsection{Optical conductivity}

\begin{figure}[t]
		\centering
	\includegraphics[width=\columnwidth]{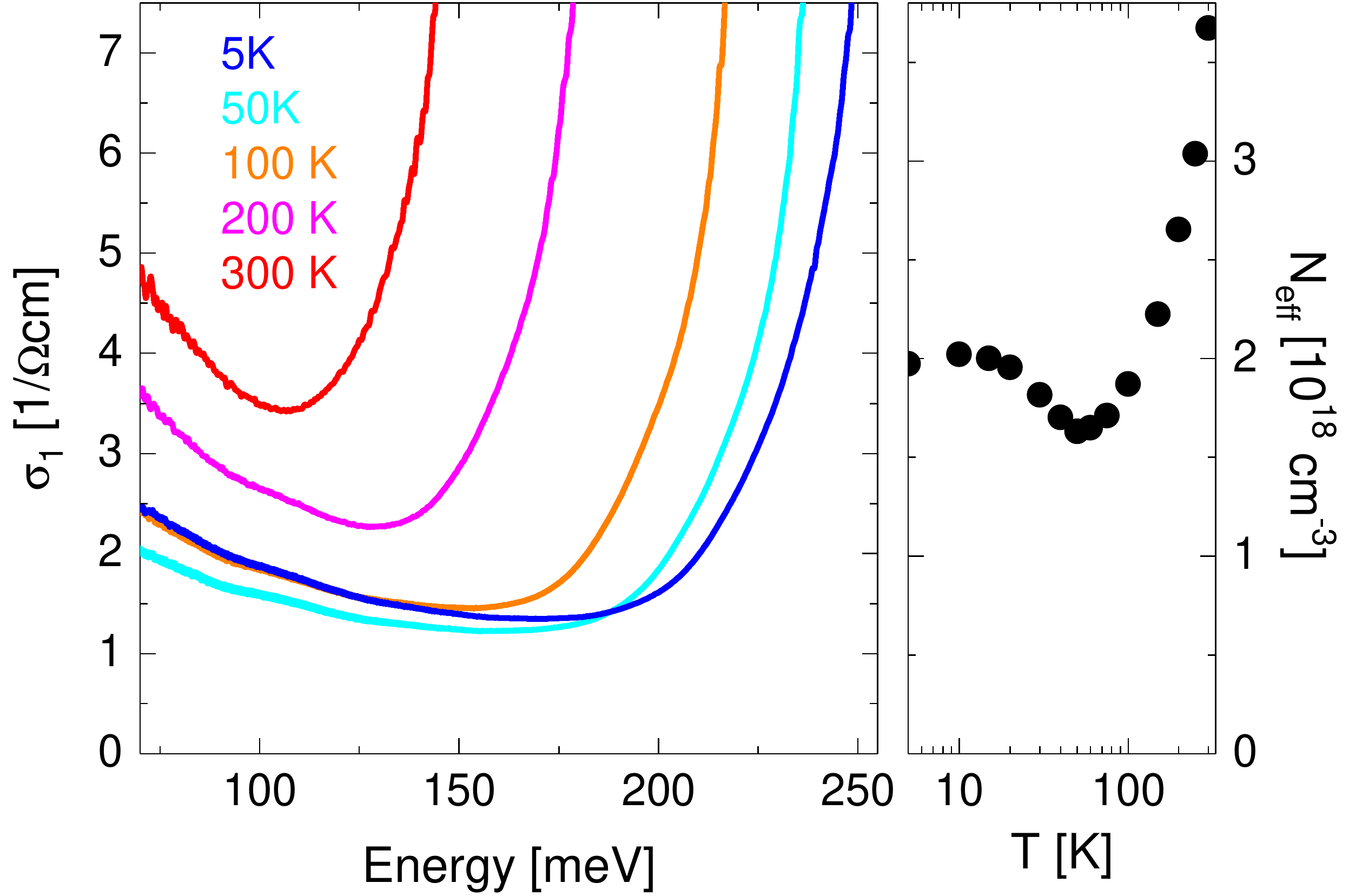}
	\caption{Left: optical conductivity $\sigma_1(\omega)$ of BiSbTeSe$_2$ in the frequency range above 
		the phonons and below the band gap, showing a broad temperature-dependent low-energy feature which we attribute 
		to a Drude-like peak. 
	    Right: effective carrier density $N_{\rm eff}$ as obtained from a Drude fit of $\sigma_1(\omega)$. 
	    The non-monotonic behavior of $N_{\rm eff}(T)$ directly tracks the non-monotonic behavior of $\sigma_1(\omega_0,T)$ 
	    at, e.g., $\hbar \omega_0$\,=\,$100$\,meV. }
	\label{fig_sigma1}
\end{figure}

Charge carriers localized in puddles do not contribute directly to the DC conductivity, but for frequencies above a 
cutoff $\omega_c$ they cannot be distinguished from free carriers. For BiSbTeSe$_2$, the estimated cutoff is very small, 
$\hbar \omega_c \leq $\,0.01\,meV \cite{Borgwardt16}. For frequencies much larger than $\omega_c$, 
bulk puddles give rise to a weak Drude-like contribution to the optical conductivity $\sigma_1(\omega)$ \cite{Borgwardt16}. 
Such weak absorption features can be determined very well in transmittance measurements. 
The frequency range in which the sample is transparent then determines the experimentally accessible range of $\sigma_1(\omega)$.  
The left panel of Fig.\ \ref{fig_sigma1} shows $\sigma_1(\omega)$ of BiSbTeSe$_2$ as obtained from a sample with a thickness 
of $d$\,=\,$(135 \pm 5)$\,$\mu$m, where $d$ was determined from the observed Fabry-Perot interference fringes. 
For $\sigma_1(\omega) \gtrsim 7/\Omega$cm, the transmittance of this particular sample is below the noise level. 
Overall, the data agree very well with our previous results on BiSbTeSe$_2$ \cite{Borgwardt16}. 
At 5\,K, the steep increase at about 0.25\,eV denotes the onset of excitations across the band gap $\Delta$. 
The broad feature at lower energy can be described in terms of a Drude peak. By fitting the spectra, we determined 
the spectral weight of this Drude peak which is a measure of the effective carrier density $N_{\rm eff}$. 

At high temperature, the Drude peak corresponds to thermally activated carriers which also contribute to the DC conductivity. 
The spectral weight of the activated Drude peak and thus $N_{\rm eff}$ drop steeply with decreasing temperature 
down to $\sim 70$\,K, 
see right panel of Fig.\ \ref{fig_sigma1}. 
The presence of puddles is revealed by the non-monotonic temperature dependence of $N_{\rm eff}$, 
i.e., the reincrease of $N_{\rm eff}(T)$ below about 50\,K.\@

By comparison with theory \cite{Borgwardt16}, the optical data allow us to estimate the carrier density $N_p$ in 
the bulk puddles, the defect density $N_{\rm def}$, and the degree of compensation $K$. 
We find $N_p$\,=\,$N_{\rm eff}\, m^*/m_e$\,=\, $4\cdot 10^{17}$\,cm$^{-3}$ using $N_{\rm eff}(5\,K)$ (see right panel 
of Fig.\ \ref{fig_sigma1}) and an estimate of the effective band mass of $m^* \! \approx \! 0.2 m_e$ \cite{Borgwardt16}, 
where $m_e$ is the free electron mass. 
For an estimate of $N_{\rm def}$, we first consider perfect compensation, $K$\,=\,$1$. 
In this case, theory predicts $N_p/N_{\rm def}$\,=\,$0.06\, E_C/\Delta$ 
with the Coulomb interaction $E_C$ between neighbouring dopants defined in Eq.\ (\ref{ec}). 
With $\varepsilon$\,=\,$200$ \cite{Borgwardt16} and our experimental result of $\Delta/k_B \approx 3000$\,K, 
we find $N_{\rm def} \approx 4\cdot 10^{20}$\,cm$^{-3}$ which is equivalent to $E_C \approx 60$\,K.\@ 
In the case of small deviations from perfect compensation such as $K$\,=\,0.98, 
theory predicts $N_p/N_{\rm def}$\,=\,$0.316\, |1-K|$. 
This yields $N_{\rm def} \approx 1\cdot 10^{20}$\,cm$^{-3}$ or $E_C \approx 40$\,K.\@ 
Our experimental data shown in the right panel of Fig.\ \ref{fig_sigma1} agree with the suppression of bulk puddles 
on a temperature scale of $E_C/k_B$\,=\,$40-60$\,K.\@ 
We thus conclude that this sample of BiSbTeSe$_2$ is close to perfect compensation and shows a defect density 
of $N_{\rm def}$\,=\,$1-4\cdot 10^{20}$\,cm$^{-3}$.

\subsection{STM and STS data}

\begin{figure}[t]
	\centering
	\includegraphics[width=8cm]{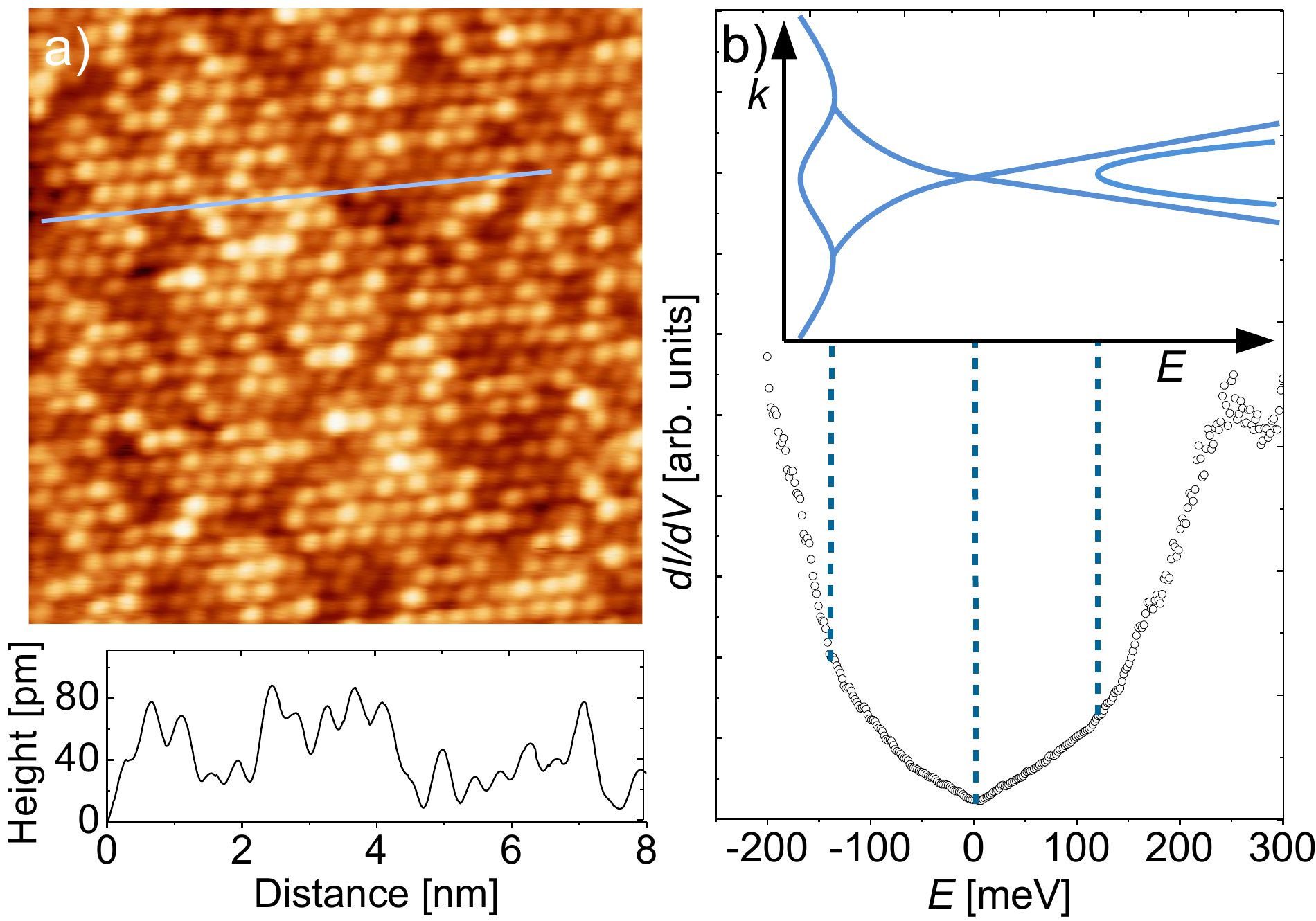}
	\caption{(a) Atomically resolved STM topograph of cleaved BiSbTeSe$_2$ ($10$\,nm\,$\times$\,$10$\,nm, $V$\,=\,$-300$\,mV, 
	$I$\,=\,$50$\,pA). The height profile along the blue line is plotted below the STM image. 
	(b) Characteristic STS spectrum at 5.5\,K.\@ The main spectral features (bulk valence band maximum, Dirac point, 
	and bulk conduction band minimum) are marked by dashed lines and linked to the sketch of the band structure in the inset, 
	which follows the ARPES result for BiSbTeSe$_2$ \cite{Arakane12}.}
	\label{surfacestructure}
\end{figure}

\begin{figure}[t]
\centering
\includegraphics[width=8cm]{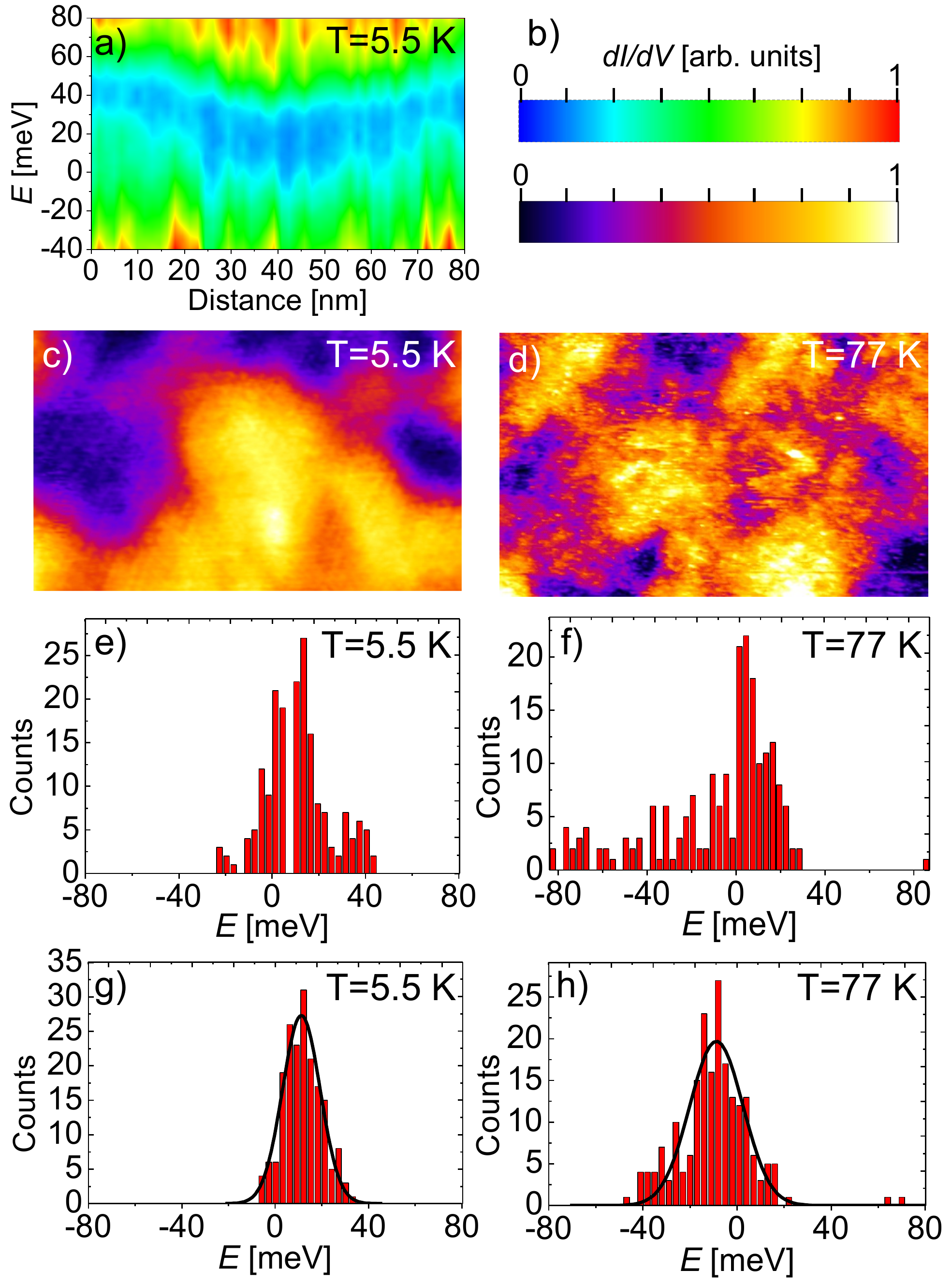}
\caption{(a) 50 STS spectra over an energy range from $-40$\,meV below the Fermi level to $80$\,meV above, 
taken in equidistant steps along a straight line of 80\,nm length.  
The shift of the Dirac point (lowest d$I$/d$V$ value) is of the order of 20\,meV.\@ 
(b) Color codes for (a), (c), and (d).
(c)  STS map recorded at 5.5\,K ($100$\,nm\,$\times$\,$60$\,nm, $V$\,=\,$50$\,mV, $I$\,=\,$70$\,pA). 
Fluctuations in the d$I$/d$V$ signal visualize the fluctuations in the energy of the Dirac point. 
(d) STS map recorded at 77\,K ($100$\,nm\,$\times$\,$60$\,nm, $V$\,=\,$80$\,mV, $I$\,=\,$50$\,pA). 
(e) Distribution of the Dirac point energy at 5.5\,K.\@ 
(f) Same distribution at 77\,K (see main text). 
(g) and (h) show the same data as in (e) and (f), but corrected for possible tip effects (see main text). 
Gaussian fits to distributions are shown as black lines. The bins in (e) to (h) have a width of 3\,meV.}
\label{surfacedos}
\end{figure}

Large-scale STM topographs of the cleaved BiSbTeSe$_2$ sample display flat terraces larger than 500\,nm that 
are separated by steps of 10\,\AA{} height, or multiples thereof, consistent with a cleavage along the van der Waals gap 
between the quintuple layers of BiSbTeSe$_2$ (compare also \cite{Kim14}).
The atomic-resolution STM topograph of Fig.\ \ref{surfacestructure}(a) shows the hexagonal lattice of the surface atoms 
together with variations in the apparent atom heights. Through the height profile along the blue line in the topograph 
of Fig.\ \ref{surfacestructure}(a), these height variations are quantified to be on the order of 50\,pm. 
The variations are interpreted to result from the random arrangement of the chemical species in the mixed topmost 
Te/Se layer and in the mixed subsurface Bi/Sb layer. The chemical inhomogeneity is the origin of the apparent variation 
of the local electronic structure on the atomic scale. 
Similar observations were reported for the sister compound Bi$_{1.5}$Sb$_{0.5}$Te$_{1.7}$Se$_{1.3}$ \cite{Kim14,Ko16}.
The cleaved surface is free of point defects (e.g.\ vacancies) and adsorbates.  

Figure \ref{surfacestructure}(b) displays a typical STS spectrum which shows the d$I$/d$V$ signal versus energy $E$\,=\,$eV$. 
The relation of features in the STS spectrum with the band structure as derived from ARPES 
\cite{Arakane12,Kim14} is visualized by the sketch in the inset. 
We find a dip in the d$I$/d$V$ data that is attributed to the Dirac energy of the surface state. The Dirac point is located 
close to the Fermi energy, in agreement with ARPES results on BiSbTeSe$_2$ \cite{Arakane12}. The top of the valence band 
is located at about -140\,meV and the bottom of the conduction band at about 125\,meV.\@ Both are characterized by a slight 
change in the slope of the d$I$/d$V$ signal. This implies a band gap of about 265\,meV, in excellent agreement with the 
optical data of the same sample discussed above, see left panel of Fig.\ \ref{fig_sigma1}. 
In the energy range within the bulk band gap, the STS spectrum results exclusively from the surface state. 
The hallmark of an ideal Dirac cone is the linear dependence of the local density of states on energy, as observed for the 
unoccupied states at positive energy. Slight deviations from perfect linearity within the gap at negative energies reflect 
the well-known curvature of the topological surface state observed in ARPES on Bi$_{2-x}$Sb$_x$Te$_{3-y}$Se$_y$ \cite{Arakane12}.

Local potential fluctuations due to ionized donors and acceptors lead to local energy shifts of the 
surface band structure and thus of the energy location of the Dirac point. 
Figure \ref{surfacedos}(a) represents these fluctuations through a sequence of 50 STS spectra 
taken along a line of 80\,nm length. 
The d$I$/d$V$ signal as a function of position and energy is visualized by a color scale ranging 
from blue to red as indicated in Fig.\ \ref{surfacedos}(b). 
The minimum in the d$I$/d$V$ signal again corresponds to the Dirac point which smoothly shifts in energy 
with the lateral coordinate, whereby the characteristic length scale as given by the distance between minimum 
and maximum energy is of the order of $r_s$\,=\,$40-50$\,nm. 

Figure \ref{surfacedos}(c) displays the corresponding d$I$/d$V$ map taken near the Dirac energy at 5.5\,K.\@ 
The change in d$I$/d$V$ [color code in Fig.\ \ref{surfacedos}(b)] corresponds to the energy shift of the surface 
band structure in line with the formation of surface puddles as shown in Fig.\ \ref{surfacedos}(a). 
Again a smooth shift of the surface band structure with respect to the Fermi level is visible, 
showing the same characteristic length scale of about $r_s$\,=\,$40-50$\,nm from minimum to maximum.

We also performed STM and STS measurements at 77\,K to investigate the temperature dependence of the puddles, 
which is significant in the bulk, see Fig. \ref{fig_sigma1}.
Figure \ref{surfacedos}(d) displays a d$I$/d$V$ map taken at 77\,K.\@ 
It is more noisy but otherwise hardly distinguishable from the ones taken at 5.5\,K.\@ 
At first sight, this is remarkable since bulk puddles evaporate on a temperature scale of $40-60$\,K, as discussed above.

To quantitatively characterize potential fluctuations, we took point spectra along lines of 80\,nm to 100\,nm length 
at 5 (7) different sample locations resulting in a total of 185 (200) spectra at 5.5\,K (77\,K). 
The results are shown in Figs.\ \ref{surfacedos}(e) and (f) for 5.5\,K and 77\,K, respectively. 
The energy of the dip found in each STS spectrum signifying the local Dirac point position $E_D$\,=\,$e \phi$ 
is extracted and collected in bins of 3\,meV, where $\phi$ denotes the local electric potential. 
The potential fluctuations can be characterized quantitatively through the standard deviation 
$\Gamma$\,=\,$e\sqrt{\frac{1}{N}\cdot\sum_{i}(\phi_i-\langle\phi\rangle)^2}$. 
At 5.5\,K the average doping level of the sample is 
$\langle E_D \rangle $\,=\,$e \langle \phi \rangle$\,=\,$11$\,meV and the magnitude of the potential fluctuations 
is $\Gamma$\,=\,14\,meV.\@ 
At 77\,K we find $\langle E_D \rangle $\,=\,$-8$\,meV and $\Gamma$\,=\,$28$\,meV.\@
This representation of the data can be considered to be an upper bound for the potential fluctuations, 
as it assumes the absence of any tip-related effects on the data. 

However, due to unavoidable occasional changes in the microscopic tip structure, e.g.\ by pick up of sample atoms, 
the tip density of states may change slightly between different locations at which the lines of spectra were taken. 
This may have an effect on the average measured doping level. 
To account for such possible parasitic effects related to the density of states of the tip, we subtracted 
the average Dirac energy of a line of spectra prior to their insertion into the histograms shown in 
Figs.\ \ref{surfacedos}(g) and (h) and centered these histograms at the respective global average.
Applying this procedure, the standard deviation of the distribution measures a lower bound of the potential fluctuations. 
The magnitude of the potential fluctuations then is $\Gamma$\,=\,$8$\,meV at 5.5\,K and $\Gamma$\,=\,$15$\,meV at 77\,K.\@ 
Here the distributions can be reasonably well fitted by Gaussians, which are shown in Figs.\ \ref{surfacedos}(g) and (h) as black lines.

Globally, it is fair to state that the samples of BiSbTeSe$_2$ are close to perfect compensation with 
the average doping level close to zero, in excellent agreement with our optical data. 
Though our estimates of $\Gamma$\,=\,$8-14$\,meV at 5.5\,K do not allow to precisely specify the 
magnitude of the potential fluctuations, we may safely conclude that $\Gamma$ is small, about an 
order of magnitude smaller than the bulk band gap. Despite our limited statistics and possible sources 
of systematic errors, we further conclude that $\Gamma$ is of similar magnitude at 5.5\,K and 77\,K.\@ 
While the fluctuations of the bulk potential decrease with increasing temperature due to thermally activated carriers, 
we find no evidence for a decrease of the magnitude $\Gamma$ of surface potential fluctuations from 5.5\,K to 77\,K.\@ 
With some reservation, our data rather suggest a slight increase.

\section{Discussion}

Using a self-consistent theory based on Thomas-Fermi screening by Dirac-like surface carriers, 
Skinner and Shklovskii  \cite{Skinner13} calculated the amplitude of potential fluctuations 
$\Gamma$\,=\,$e \langle (\phi-\langle\phi\rangle)^2 \rangle^{1/2}$ at the TI surface as well as 
the characteristic size $r_s$ of surface puddles caused by Coulomb disorder in the bulk 
for perfect compensation, $N_{\rm def}$\,=\,$N_D$\,=\,$N_A$. For nominally undoped TIs in which 
the Dirac point lies at the chemical potential, $E_D$\,=\,$0$, they find 
\begin{equation}
r_s = (4\, \alpha_{\rm eff}^4 N_{\rm def})^{-1/3}
\label{eq_rs}
\end{equation}
with the \textit{effective} fine structure constant 
\begin{equation}
\alpha_{\rm eff} = \frac{e^2}{4\pi \varepsilon_0 \varepsilon_{\rm eff} \hbar v_F} 
= \alpha \,\frac{c}{\varepsilon_{\rm eff} v_F} \, ,
\end{equation}
where $\alpha$\,=\,$1/137$ denotes the fine structure constant, 
$c$ the speed of light, and $v_F$ the Fermi velocity. 
The effective dielectric constant at the surface is given by $\varepsilon_{\rm eff}$\,=\,$(\varepsilon+1)/2$, 
where $\varepsilon \approx 200$ \cite{Borgwardt16} denotes the bulk dielectric constant. 
In Bi$_{2-x}$Sb$_x$Te$_{3-y}$Se$_y$, the dispersion of the topological surface state is not perfectly linear, 
giving rise to a variation of $v_F$ in the range of $3-5 \cdot 10^5$\,m/s \cite{Arakane12,Neupane12,Kim14,Ko16}. 
Altogether, we find $\alpha_{\rm eff}$\,=\,$(8\pm2)\,\alpha  \approx 0.04-0.07$, significantly smaller than 
the value of 0.24 estimated by Skinner and Shklovskii for Bi$_2$Se$_3$ \cite{Skinner13}. 
With the optically determined defect density $N_{\rm def}$\,=\,$1-4 \cdot 10^{20}$\,cm$^{-3}$, 
we find $r_s \! \approx \! 30-90$\,nm, in very good agreement with our STS result of $40-50$\,nm.

Furthermore, Skinner and Shklovskii \cite{Skinner13} find
\begin{equation}
\Gamma = 4 \frac{E_C}{\alpha_{\rm eff}^{2/3}} 
\,\, \propto \,\, \left( \frac{v_F^2 N_{\rm def}}{\varepsilon}\right)^{1/3} \, ,
\label{eq_Gamma}
\end{equation}
again for nominally undoped samples, $E_D$\,=\,$0$. 
With our experimental result of $E_C/k_B$\,=\,$40-60$\,K derived from the optical data, 
this yields a theoretical prediction of $\Gamma$\,=\,$80-170$\,meV, much larger than the 
value of $8-14$\,meV observed in STS at 5.5\,K.\@ 
In the following we show that this inconsistency between theory and experiment can also be 
derived using only the STS data, without reference to our optical results. 
Theory predicts \cite{Skinner13}
\begin{equation}
r_s \, \Gamma  = \frac{4^{2/3} \, E_C}{\alpha_{\rm eff}^2 N_{\rm def}^{1/3}} = 
4^{2/3} \pi \varepsilon_0 (\hbar/e)^2 \cdot \varepsilon v_F^2
\label{eq_Gammars}
\end{equation}
which depends only on $\varepsilon$ and $v_F$. The resulting prediction of $r_s \, \Gamma$\,=\,$3-9$\,eVnm 
is an order of magnitude larger than the experimental STS result of $0.3-0.7$\,eVnm at 5.5\,K.\@

Equations (\ref{eq_rs}) and (\ref{eq_Gamma}) were derived for nominally undoped samples but are expected 
to apply as long as the chemical potential $\mu$ is smaller or at least not much larger than the potential fluctuations $\Gamma$. 
For doped compounds with $|\mu| \gg 2E_C/\alpha_{\rm eff}^{2/3}$ [cf.\ Eq.\ (\ref{eq_Gamma})], 
Skinner and Shklovskii find \cite{Skinner13}
\begin{equation}
r_s = \frac{\hbar v_F}{\alpha_{\rm eff}\, |\mu|} \,\,\,\,\,\,\,\, {\rm and} \,\,\,\,\,\,\,\, 
\Gamma^2  = \frac{16\pi\, E_C^3}{\alpha_{\rm eff}^2|\mu|} \, .
\label{eq_rsmue}
\end{equation}
This limit of large $\mu$ is not applicable in BiSbTeSe$_2$, as demonstrated by our STS results. 
In fact, Eq.\ (\ref{eq_rsmue}) predicts an even larger value of $\Gamma$ than Eq.\ (\ref{eq_Gamma}) 
for the experimentally determined parameters.

The calculations of Skinner \textit{et al.} \cite{Skinner12,Skinner13} neglect the screening contribution 
of bulk puddles, i.e., they assume that all donors and acceptors are ionized. 
In other words, they assume that the formation of surface puddles is \textit{not} affected 
by bulk puddles. This is valid for $r_s \ll l_s$, where $l_s$ denotes the characteristic thickness of the 
near-surface layer in which bulk puddles are suppressed. 
For $r_s \ll l_s$, the Coulomb disorder near the surface is screened by surface carriers with potential 
fluctuations $\Gamma$ (much) smaller than $\Delta/2$, preventing strong band bending and the formation of 
bulk puddles. 
Below, we will argue that in our system the assumption $r_s \ll l_s$ is not valid.

The size $R$ of bulk puddles -- not to be confused with the thickness $l_s$ of the near-surface layer -- 
was estimated by Shklovskii and coworkers by a simple scaling argument \cite{Skinner12}. 
It states that random fluctuations of the density of ionized defects in a volume $R^3$ give rise to 
an uncompensated charge $\propto (N_{\rm def}R^3)^{1/2}$ and thus to a Coulomb potential $\propto \sqrt{R}$. 
Bulk puddles are formed if these potential fluctuations are as large as $\Delta/2$, which leads to the estimate 
\cite{Skinner12}
\begin{equation}
R = \frac{(\Delta/E_C)^2}{8\pi \, N_{\rm def}^{1/3}} \,.
\label{eq_R}
\end{equation}
Using $\Delta$\,=\,0.25\,eV, $E_C$\,=\,$40-60$\,K, and $N_{\rm def}$\,=\,$1-4\cdot 10^{20}$\,cm$^{-3}$, 
we find $R \approx 100-500$\,nm. However, recent numerical results combined with a refined scaling argument 
by B\"{o}merich \textit{et al.} \cite{Boemerich17} predict much smaller values 
with $R\propto (\Delta/E_C)^{1.1}$ for $\Delta/E_C \leq 40$ and
\begin{equation}
R \propto \frac{(\Delta/E_C)^2}{\ln(\Delta/E_C)} 
\label{eq_Rlog}
\end{equation}
for $\Delta/E_C \to \infty$. 
The same behavior but with a smaller prefactor is found for $l_s$ \cite{Boemerich17}. 
Under the assumption that the surface carriers screen like a perfect metal, i.e., large $|\mu|$, 
B\"{o}merich \textit{et al.} predict $l_s \approx 7-12 /N_{\rm def}^{1/3}$ in the range relevant to us, 
i.e., $\Delta/E_C$\,=\,$50-75$. This corresponds to a thickness of the near-surface layer $l_s$\,=\,$9-25$\,nm 
which is even \textit{smaller} than the measured characteristic size of surface puddles $r_s$\,=\,$40-50$\,nm.
The assumption of perfectly metallic surfaces with a high chemical potential rather overestimates 
the value of $l_s$ in BiSbTeSe$_2$. This result thus clearly suggests that the assumption $r_s \ll l_s$ 
is not valid anymore. Therefore, screening by bulk puddles may contribute to the surface properties, 
effectively reducing the amplitude of potential fluctuations on the surface. 
Altogether, these results indicate that surface puddles are not fully independent from bulk puddles and 
that a full quantitative description of the experimental data requires to consider a self-consistent description 
of both, surface and bulk properties. 

Finally, we address the temperature dependence. The optical data show that bulk puddles "evaporate" with 
increasing temperature, the temperature scale is given by $E_C/k_B$\,=\,$40-60$\,K in our sample of BiSbTeSe$_2$. 
In contrast, no strong change of the surface properties is observed in STS between 5.5\,K and 77\,K.\@ 
Naively, this seems to suggest that surface puddles are independent of bulk carriers. 
However, the suppression of bulk puddles is due to the thermal activation of carriers in the bulk. 
Accordingly, the screening capability of the bulk is smoothly enhanced with increasing temperature. 
Bulk puddles vanish because the activated carriers reduce the potential fluctuations to a value smaller than 
$\Delta/2$. In other words, screening in the bulk at low temperatures is due to bulk puddles, but this role 
is taken over by activated carriers with increasing temperature. Therefore, it is very well possible that 
the temperature scale for the suppression of bulk puddles has little relevance to surface puddles.

\section{Summary and Conclusions}

To address the relationship of surface puddles and bulk puddles, we performed infrared 
transmittance, STM, and STS investigations of the same sample of BiSbTeSe$_2$. 
Measurements on the same sample are important since essential parameters such as the 
defect density $N_{\rm def}$ and the degree of compensation $K$ are sample dependent. 
By \textit{in-situ} cleaving, large-scale flat terraces of more than 500\,nm were obtained. 
In agreement with previous results, STM and STS data show the hexagonal atomic structure 
and the chemical inhomogeneity of the Te/Se layers \cite{Kim14,Ko16}. 
Both optics and STS find a band gap of about 0.25\,eV and show that BiSbTeSe$_2$ is nearly undoped 
and close to perfect compensation. 

Our data demonstrate the coexistence of bulk puddles and surface puddles at low temperatures. 
Both are caused by the Coulomb disorder originating from randomly distributed, ionized donors 
and acceptors. Although sharing the same origin, their properties are very different. 
Bulk puddles contribute to screening by neutralizing donors and acceptors, while the screening 
by surface puddles is based on the redistribution of highly mobile Dirac-like metallic carriers. 
Accordingly, the formation of bulk puddles requires that the potential fluctuations are as large 
as $\Delta/2$ such that the bands touch the chemical potential. In contrast, surface puddles are 
formed for any finite variation of the surface potential since this will give rise to local shifts 
of the chemical potential for the gapless surface state. 
This major difference in energy scales explains their entirely different behavior as a 
function of temperature as well as the different length scales of surface puddles and bulk puddles. 
Bulk puddles "evaporate" at a temperature scale of $E_C$ when the screening contribution of 
activated carriers reduces the amplitude of potential fluctuations below $\Delta/2$. 
This temperature scale has little relevance to surface puddles since the \textit{total} screening 
properties of the bulk evolve smoothly with temperature. 

The quantitative analysis of our data allows us to address a possible interrelation of 
surface puddles and bulk puddles.
Our optical data yield $K$\,=\,$0.98-1$, $N_{\rm def}$\,=\,$1-4 \cdot 10^{20}$\,cm$^{-3}$, 
and a value of $E_C/k_B$\,=\,$40-60$\,K for the average Coulomb interaction 
between neighboring dopants.\@ 
At 5.5\,K, the STS measurements reveal an amplitude $\Gamma$\,=\,$8-14$\,meV of the 
surface potential fluctuations which occur on a length scale of $r_s$\,=\,$40-50$\,nm. 
The value of $\Gamma$ is much smaller than $\Delta/2$, which is the size of potential fluctuations 
in the bulk.
We quantitatively compared the experimental results with the predictions of a self-consistent theory 
based on Thomas-Fermi screening by Dirac-like carriers \cite{Skinner13} which neglects the 
screening by bulk carriers. This is valid under the assumption $r_s \ll l_s$, i.e., the length 
scale of surface puddles is much smaller than the thickness of the near-surface layer in which 
bulk puddles are suppressed. 
Our experimental result for $\Gamma$ is about an order of magnitude smaller 
than predicted by theory. In other words, screening by Dirac-like carriers is not sufficient 
to explain the small magnitude $\Gamma$ of surface potential fluctuations 
in nearly undoped BiSbTeSe$_2$ with a small value of the chemical potential.
However, numerical results in combination with a refined scaling argument reported recently 
\cite{Boemerich17} indicate that the assumption $r_s \ll l_s$ does not hold in BiSbTeSe$_2$. 
We therefore conclude that surface puddles most probably are not fully independent from 
bulk puddles in BiSbTeSe$_2$, i.e., the screening contribution of the bulk is relevant for 
surface properties at least as long as the surface chemical potential is close to the Dirac point.
Further theoretical studies are called for to clarify whether quantitative agreement between 
experiment and theory indeed requires a self-consistent description of both, 
surface and bulk properties at the same time.

\begin{acknowledgments}
	Financial support of the Deutsche Forschungsgemeinschaft (DFG)
	through the Collaborative Research Center SFB 1238 (projects A04, B02, B06, and C02)
	is gratefully acknowledged. W.J. acknowledges financial support by 
	the University of Cologne via the Advanced Postdoc Grant '2D materials beyond graphene'
	(PI: C. Busse).
\end{acknowledgments}

\end{document}